\documentclass{article}
\RequirePackage{graphicx}
\RequirePackage{amssymb,amsmath,bm}
\RequirePackage{textcomp}
\RequirePackage{booktabs}
\RequirePackage[textfont=it,tableposition=top]{caption}
\RequirePackage{siunitx}
\RequirePackage{xspace}
\RequirePackage{url}
\RequirePackage{lipsum}  
\RequirePackage{tipa}
\RequirePackage{enumitem}
\RequirePackage{xkeyval}
\RequirePackage{calc}

\usepackage{spconf,amsmath,graphicx}
\usepackage{amsmath}
\usepackage{multirow}
\usepackage{hyperref}
\usepackage[utf8]{inputenc}

\title{Addressing Index Collapse of Large-Codebook Speech Tokenizer with Dual-Decoding Product-Quantized Variational Auto-Encoder}

\name{Haohan Guo$^*$,
    Fenglong Xie$^\dag$,
    Dongchao Yang$^*$,
    Hui Lu$^*$,
    Xixin Wu$^*$,
    Helen Meng$^*$}
\address{
    $^*$The Chinese University of Hong Kong, Hong Kong SAR, China \\
    $^\dag$Xiaohongshu Inc., Shanghai, China \\
    \href{mailto:hguo@se.cuhk.edu.hk}{\nolinkurl{{hguo, dcyang, luhui, xxwu, hmmeng}@se.cuhk.edu.hk}}, \\
    \href{mailto:fenglongxie@xiaohongshu.com}{{\nolinkurl{fenglongxie@xiaohongshu.com}}}
}

\begin{document}

\maketitle

\begin{abstract}

VQ-VAE, as a mainstream approach of speech tokenizer, has been troubled by ``index collapse'', where only a small number of codewords are activated in large codebooks. This work proposes product-quantized (PQ) VAE with more codebooks but fewer codewords to address this problem and build large-codebook speech tokenizers. It encodes speech features into multiple VQ subspaces and composes them into codewords in a larger codebook. Besides, to utilize each VQ subspace well, we also enhance PQ-VAE via a dual-decoding training strategy with the encoding and quantized sequences. The experimental results demonstrate that PQ-VAE addresses ``index collapse" effectively, especially for larger codebooks. The model with the proposed training strategy further improves codebook perplexity and reconstruction quality, outperforming other multi-codebook VQ approaches. Finally, PQ-VAE demonstrates its effectiveness in language-model-based TTS, supporting higher-quality speech generation with larger codebooks.
    
\end{abstract}

\section{Introduction}

The large language model (LLM) has demonstrated its powerful capability in text generation \cite{brown2020language,openai2023gpt4,touvron2023llama2}. It can auto-regressively generate expressive and diverse text sequences, especially when scaled to a larger model with more training data \cite{kaplan2020scaling}. This breakthrough has also attracted widespread attention from the speech domain, inspiring the next-gen speech generation paradigm, i.e. speech language model \cite{hassid2024textually,wu2023speechgen}. To apply LLM to the speech domain, the first thing is to convert the long speech sequence based on continuous representations into a short sequence with discrete tokens to mimic the text. Then, we can combine speech and text together for LLM training and inference, and achieve conditional speech generation, e.g. text-to-speech (TTS) \cite{betker2023better,lajszczak2024base}, voice conversion (VC) \cite{wang2023lm,kuan2023towards}, and speech-to-speech translation \cite{huang2023speech}. Hence, a high-quality speech tokenizer providing discrete tokens with sufficient speech information is the key to generating intelligible and natural speech.

Vector-quantized variational autoencoder (VQ-VAE) \cite{vqvae}, as the mainstream model for discrete representation learning of speech, has been well applied in multiple tasks, including speech coding \cite{garbacea2019low}, VC \cite{wang2021vqmivc}, and TTS \cite{Du2022VQTTSHT}. It also demonstrates great potential in speech tokenization over conventional k-means-based approaches \cite{huang2023repcodec}. However, in practice, training a high-quality VQ-VAE-based speech tokenizer is also challenging due to the problem of ``index collapse" \cite{huh2023improvedvqste}. The speech tokenizer usually needs a large codebook comparable to the text dictionary to represent rich speech information. VQ-VAE with a large codebook often fails to learn all codewords well, resulting in only a small number of codewords being activated in both training and inference. Some works turn to exploit multi-sequence discrete representations, e.g. MSMC-VQ \cite{guo2023msmc} and RVQ \cite{DBLP:journals/corr/abs-2210-13438,wang2023neural,zhang2023speechtokenizer}, to avoid learning large codebooks, but they cannot adapt LLMs only modeling single discrete sequences directly, introducing more challenges.


To avoid this dilemma, we aim to address the ``index collapse" of VQ-VAE to support the training of large-codebook speech tokenizers. We propose PQ-VAE, which indirectly learns a large codebook from multiple small codebooks by replacing VQ with product quantization (PQ) \cite{jegou2010product}. Besides, we enhance the training process with the proposed dual-decoding training strategy to pursue a higher-quality codebook. We conduct experiments on a large-scale speech dataset, first revealing the phenomenon of ``index collapse" when the codebook is enlarged and then demonstrating the effectiveness of PQ-VAE in addressing this problem. Finally, we investigate the performance of PQ-VAE in TTS to show further the importance of a large-codebook speech tokenizer without ``index collapse" for LLM-based speech generation.

\section{Problem Statement}

\subsection{Quick Review of VQ-VAE}

First, we review the vector-quantized variational autoencoder (VQ-VAE). This model comprises an encoder, a decoder, and a quantizer. The encoder first processes the input vector $x$ to an encoding vector $e$. Then, in the quantizer, we replace $e$ with the closest codeword in the codebook $C = \{c_0, c_1, ..., c_{N-1}\}$ with the size of $N$ as the quantized vector $z$, i.e.
\begin{equation}
    z = c_{i^*} \ \ where \ \ i^* = \mathop{\arg \min}_{0 \le i < N} ||e - c_i||^2_2
\end{equation}
where $i*$ is the discrete representation (i.e. index) of $x$. Finally, we feed $z$ to the decoder to obtain the reconstructed vector $\hat{x}$. The model is trained with the loss function:
\begin{align}
    \mathcal{L} = ||x - \hat{x}||^2_2 + \alpha \ ||e - sg(z)||^2_2 + \beta \ ||sg(e) - z||^2_2
\end{align}
where $sg$ denotes ``stop gradient", $\alpha$ and $\beta$ are loss weights. The third term is usually replaced with the exponential moving average \cite{Oord2017NeuralDR} (EMA) to pursue a stable codebook update.

\subsection{Index Collapse}

``Index collapse" means that only a small fraction of codewords are activated in training, and the rest are ``dead'', which will never be used at inference. It causes low codebook usage: $|C|_u = \sum^{N-1}_{i=0} \textbf{1}(c_i) \ll N$, and perplexity: $|C|_p = 2^{-\sum^{N-1}_{i=0} p_i*\log_2p_i} \ll N$,
where $\textbf{1}(c_i)$ is 1 if $c_i$ is used, otherwise it is 0, and $p_i$ denotes the probability of $c_i$ in $C$.

This problem becomes more serious as the codebook enlarges, obstructing training large-codebook speech tokenizers. There have been some studies on this issue, and one important reason is identified: the sparse gradient of the codebook \cite{huh2023improvedvqste}. Each codeword is updated in training from only embedding vectors in the same cluster. Hence, the more vectors the codeword receives, the more accurately it updates. However, due to the constantly changing encoding space and high-variance batched data, some codewords may not receive sufficient embeddings for correct updates, and lose more in the following training, leading to death eventually. Some works propose activating dead codewords via some replacement policies \cite{huh2023improvedvqste} or controlling the distribution of codewords \cite{zheng2023online}. However, the improvement is still limited as codewords keep increasing. Hence, to avoid ``index collapse", one intuitive idea is to reduce codewords to ensure stable updates. We may use more small codebooks to learn one large codebook indirectly for the speech tokenizer, and product quantization makes it possible.

\begin{figure}[htp]
  \centering
  \includegraphics[width=\linewidth]{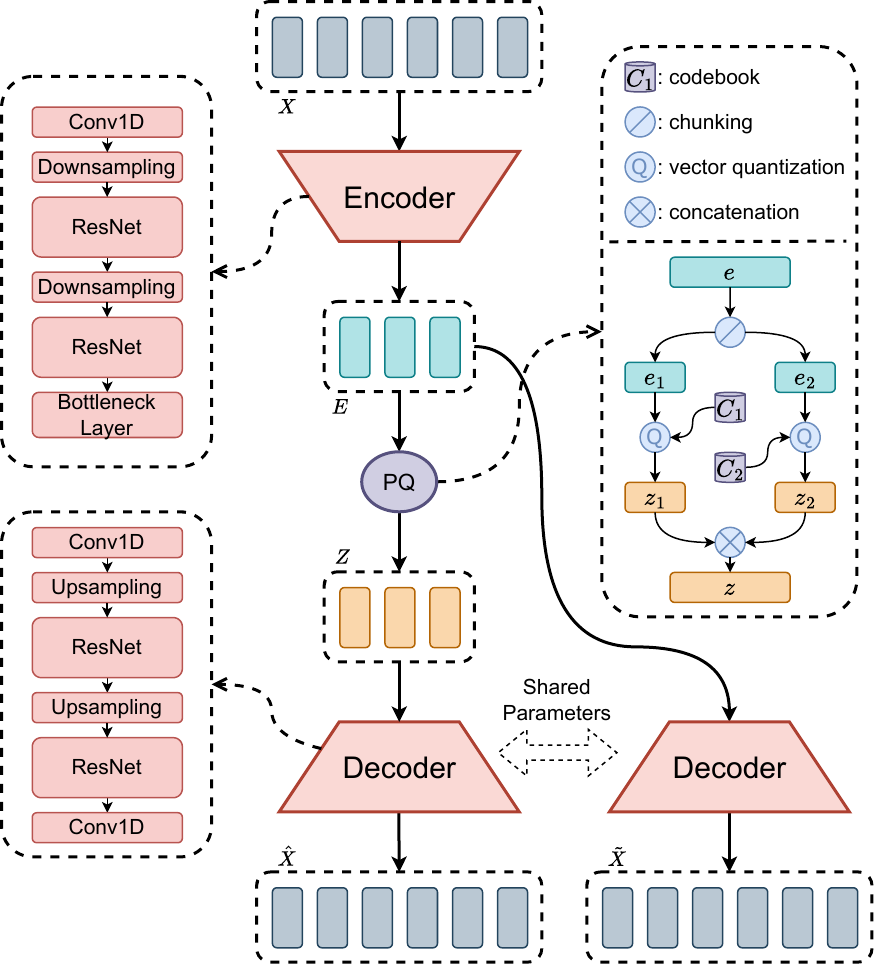}
  \caption{The framework of PQ-VAE with dual-decoding training strategy.}
  \label{fig:pqvae}
\end{figure}

\section{Dual-Decoding PQ-VAE}

As shown in Fig. \ref{fig:pqvae}, this section introduces the proposed speech tokenizer based on PQ-VAE, including the model architecture, product quantization, and the dual-decoding training strategy.

\subsection{Model Architecture}

The encoder embeds and down-samples the speech sequence $X$ to a shorter sequence $E$ using a stack of 1-D convolutional (Conv1D) layers. The down-sampling layer is a strided Conv1D layer. The ResNet comprises two residual units, where each unit has one Conv1d layer followed by two residual Conv1D layers with an exponential linear unit (ELU) in between. The bottneck layer is a linear layer, mapping the sequence into lower-dimensional vectors. After obtaining the quantized sequence $Z$ via PQ, the decoder with a similar structure as the encoder outputs the reconstructed speech sequence $\hat{X}$, where the up-sampling operation is a transposed 1-D convolutional layer. Besides, we also apply dual-decoding by decoding the encoding sequence $E$ to the speech sequence $\tilde{X}$.

\begin{figure*}[htp]
  \centering
  \includegraphics[width=\linewidth]{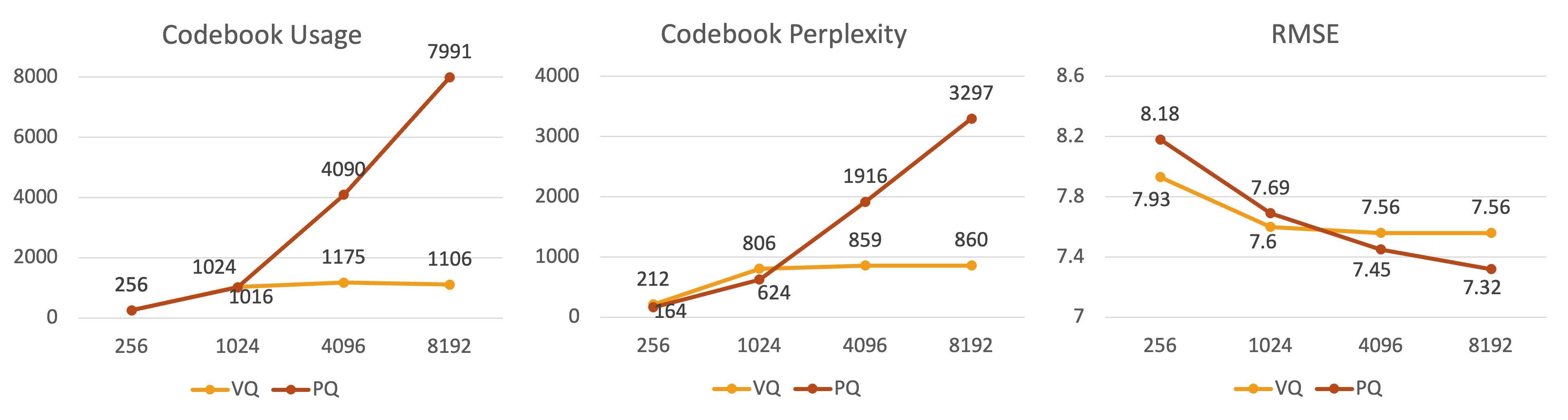}
  \caption{The codebook usage, codebook perplexity, and reconstruction loss (RMSE) of VQ-VAE and PQ-VAE under different total codebook sizes (the horizontal axis).}
  \label{fig:pqvq}
\end{figure*}

\subsection{Product Quantization}

PQ can be seen as a group of VQ modules, comprising codebooks $\textbf{C} = \{C_0, C_1, ..., C_{M-1}\}$ with the size of $M$, where the $i$-th codebook $C_i$ has $N_i$ codewords. Once an embedding $e$ comes in, we first split it into several chunks $\{e_0, e_1, ..., e_{M-1}\}$, where each sub-vector $e_i$ is quantized by the corresponding codebook $C_i$ with the same feature dimension. Finally, we concatenate these quantized sub-vectors to form the PQ output vector $z$.

PQ-VAE can be trained directly with the same loss function as the vanilla VQ-VAE, except for applying EMA updating to multiple codebooks. After training, we can create a new codebook $C^*$ by composing all well-updated codebooks in PQ together in the following way:
\begin{align}
\begin{split}    
    c^*_{i^*} &= c_{0, i_0} \otimes c_{1, i_1} \otimes \dots \otimes c_{M-1, i_{M-1}} \\
    i^* &= i_0 + \sum^{M-1}_{j=1} (\prod_{k=0}^{j-1} N_k) * i_j
\end{split}
\end{align}
where $c_{j, i_j}$ is the $i_j$-th codeword in the $j$-th codebook, $\otimes$ denotes the concatenating operation. In this way, we have a larger codebook with the size of $|C^*| = \prod_{i=0}^{M-1} N_i$. For example, we can easily create 65,536 codewords from four small codebooks with sizes of $[16, 16, 16, 16]$.

\subsection{Dual-Decoding Training Strategy}

Different from the vanilla VQ-VAE, the upper limit of codebook usage and perplexity of PQ-VAE is determined by the quality of each sub-codebook, i.e. $|C^*|_u <= \prod_{i=0}^{M-1} |C_i|_u <= |C^*|$, and $|C^*|_p <= \prod_{i=0}^{M-1} |C_i|_p <= |C^*|$. Hence, training the model to utilize each subspace equally well to produce high-perplexity codebooks is crucial. In our work, we propose first bottlenecking each subspace via a low-dimensional linear layer, then decoding both encoding sequence $E$ and quantized sequence $Z$ to reconstructed speech $\tilde{X}$ and $\hat{X}$. The loss $||X, \tilde{X}||^2_2$ encourages each subspace to carry effective information for speech reconstruction. The loss function is as follows:
\begin{align}
\label{eq:vqvae_loss}
\begin{split}
    \mathcal{L} &= ||X - \hat{X}||^2_2 + \lambda ||X - \tilde{X}||^2_2 \\
    & + \alpha * ||E - sg(Z)||^2_2 + \beta * ||sg(E) - Z||^2_2
\end{split}
\end{align}
Finally, to maximize the reconstruction quality from $Z$, we also linearly decay $\lambda$ to 0 in the late stage of training.

\begin{figure*}[htp]
  \centering
  \includegraphics[width=\linewidth]{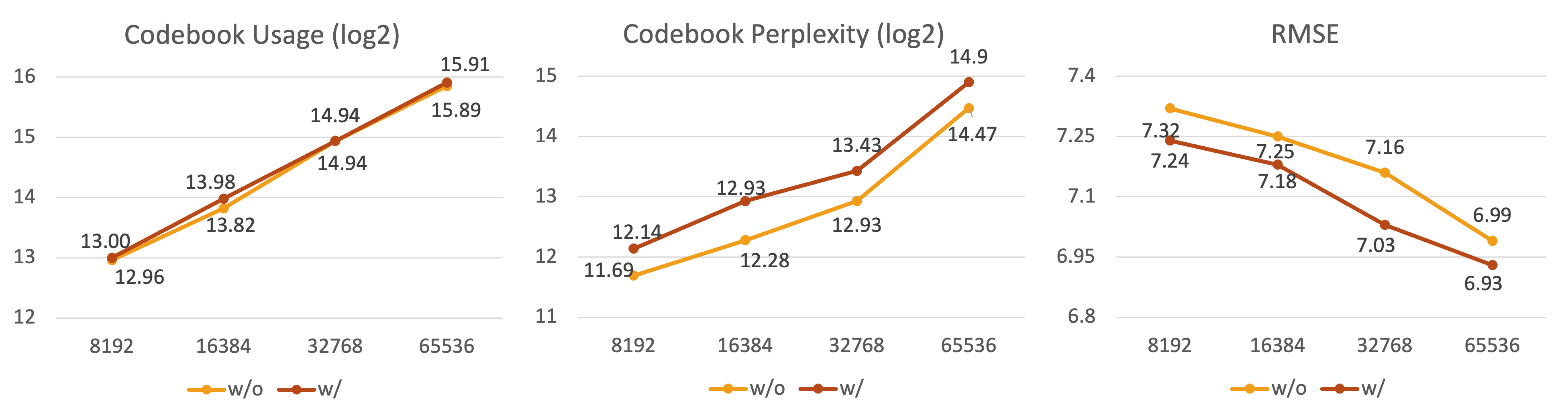}
  \caption{The codebook usage and perplexity in $log_2$ scale, and reconstruction loss (RMSE) of PQ-VAE with or without the proposed dual-decoding training strategy under different total codebook sizes (the horizontal axis).}
  \label{fig:pqdd}
\end{figure*}

\section{Experiments}

\subsection{Experimental Setup}

\subsubsection{Dataset}

We conduct experiments on a challenging large-scale dataset, WenetSpeech \cite{zhang2022wenetspeech}, containing 10,000 hours of Mandarin speech with high diversity in content, speaking style, timbre, and recording environments. We pre-process audio by downsampling to the sample rate of 16kHz and denoising \footnote{The tool is available at \url{https://github.com/Rikorose/DeepFilterNet}}. The audio that is too short ($<$ 2 seconds) or too long ($>$ 20 seconds) is removed from the training. Finally, around 7,000 hours of data are used in training.

\subsubsection{Training}

We mainly build speech tokenizers using the 80-dim log-scale Mel spectrogram with a frameshift of 10ms. The autoencoders are implemented based on 512-dim ResNet blocks with strided and transposed convolutional layers.\footnote{We implement autoencoders by following the code: \url{https://github.com/wesbz/SoundStream/blob/main/net.py}} They first downsample Mel spectrograms by 4 times to obtain $Z$ with a frameshift of 40ms, and reconstruct them back. Models are trained using AdamW \cite{loshchilov2017decoupled} ($\beta_1=0.9, \beta_2=0.95$) for 100,000 iterations with a dynamic batch size of 512 seconds. We set $\alpha = 1$ in the loss function, and use EMA to update codebooks with the decay rate of 0.999 for VQ-VAE and 0.9 for PQ-VAE. In dual-decoding training, $\lambda$ linearly decreases from 1 to 0.1 within 60k iterations starting from step 20k.

We also train LM-based TTS systems using 1024-dim HuBERT\footnote{HuBERT is available at \url{https://huggingface.co/TencentGameMate/chinese-hubert-large}} \cite{hsu2021hubert} features, a self-supervised learning (SSL) representation widely used in TTS \cite{guo2023qs}. Each speech tokenizer still compresses HuBERT features into discrete tokens with a frameshift of 40ms, and is applied with a Tortoise-style language model \cite{tortoise} with 12 512-dim transformer layers as the acoustic model, trained with a batch size of 12,800 seconds for 50,000 iterations. Finally, we use a pre-trained HuBERT-based BigVGAN-large \cite{bigvgan} vocoder for waveform generation. In TTS, we feed the Chinese characters to the language model to generate speech tokens, and feed them to the PQ-VAE decoder to obtain HuBERT features, which are then reconstructed to the audio via the vocoder.

\subsubsection{Evaluation}

We evaluate speech tokenizers with the ``Test-Net'' set in WenetSpeech, including 23 hours of audio across various domains, containing rich acoustic information. The evaluation metrics include codebook usage, perplexity, and root mean squared error (RMSE) between $X$ and $\hat{X}$. We collect 100 sentences for TTS to synthesize audio from two clean speakers with 20-second reference audio. Then, we measure the audio quality, intelligibility\footnote{The ASR tool is available at \url{https://huggingface.co/openai/whisper-large-v3}.} and speaker similarity (SS)\footnote{The speaker model is available at \url{https://huggingface.co/speechbrain/spkrec-ecapa-voxceleb}.} of the synthesized audio by calculating: 1) NISQA-TTS score \cite{mittag2021deep}, 2) character error rate (CER) between synthesized audio and ground-truth transcripts, 3) cosine distance between speaker embeddings from the synthesized audio and the reference audio.

\subsection{Index Collapse}

We first compare VQ-VAE and PQ-VAE under different codebook sizes: 256, 1024, 4096, 8192, where the codebook sizes of PQ-VAE models are assigned: $[16, 16]$, $[16, 8, 8]$, $[16, 16, 16]$, $[16, 8, 8, 8]$, and dual-decoding training is not used here for a fair comparison. The embedding has a dimension of 256, and is evenly divided to each subspace for PQ. As shown in Fig. \ref{fig:pqvq}, VQ-VAE performs well under the codebook sizes of 256 and 1024, with fully used codebooks, showing lower reconstruction loss than PQ-VAE, demonstrating its effectiveness in building the speech tokenizer with a minor or moderate codebook size. However, ``index collapse" happens when the codebook size increases to 4096 and 8192, leading to limited usage and perplexity. Meanwhile, PQ-VAE performs better as the codebook is enlarged, still achieving 98\% codebook usage for 8192 codewords, further reducing reconstruction loss. It strongly validates that PQ-VAE can address ``index collapse" effectively.

\subsection{Dual-Decoding Training}

As shown in Fig. \ref{fig:pqdd}, we further investigate the performance of PQ-VAE in building larger-codebook speech tokenizers when applied with the proposed dual-decoding training strategy. The codebook sizes are set to $[16, 16, 8, 8]$, $[16, 16, 16, 8]$, and $[16, 16, 16, 16]$, for the total codebook sizes of 16384, 32768, and 65536, respectively. As the codewords increase, PQ-VAEs with dual-decoding training keep the same high codebook usage as the vanilla PQ-VAE model, but improve the perplexity effectively to make more codewords be used more frequently. It leads to a higher reconstruction quality across different sizes of codewords, demonstrating that dual-decoding training enables PQ-VAE to become a better speech tokenizer.

We also conducted an ablation study to investigate the impact of dual-decoding training techniques on the model. As shown in Table \ref{tab:ab_study}, the vanilla PQ-VAE with 8192 codewords has already been improved in all aspects by dual-decoding training without the decayed loss and bottlenecking. The decayed weight of loss enables the model to reconstruct better speech from the quantized sequence. Finally, we obtain the lowest reconstruction loss by further bottlenecking each subspace from 64-dim to 16-dim. But we also notice that over-bottlenecking, e.g. 1-dim embedding, may degrade the model performance instead due to serious information loss.

\begin{table}[htp]
\centering
\caption{The evaluation results on test-large of PQ-VAE models.}
\label{tab:ab_study}
\begin{tabular}{l|ccc}
\toprule
Method               & Usage & Perplexity & RMSE \\ \midrule
PQ-VAE             & 7991 & 3297 & 7.32 \\
 + dual-decoding   & \textbf{8192} & 4115 & 7.28 \\ 
 + loss decay      & 8184 & 4032 & 7.26 \\
 + bottleneck      & 8190 & \textbf{4512} & \textbf{7.24} \\ \bottomrule

\end{tabular}
\end{table}

\subsection{Multi-Codebook Vector Quantization}

We also compare PQ-VAE with other multi-codebook VQ techniques: residual-vector quantization (RVQ) \cite{chen2010approximate} and finite scalar quantization (FSQ) \cite{mentzer2023finite}. RVQ quantizes a vector recursively by taking the residual vector of VQ to the next codebook for further quantization, and so on. FSQ can be seen as a special PQ with fixed codebooks, which maps speech into 1-dim subspaces, and quantizes each 1-dim vector, i.e. scalar, with pre-designed codebooks based on integers. The embedding is also constrained with a Tanh function to keep a fixed numerical range. As shown in Table \ref{tab:fsq}, we evaluate the performance of these VQ approaches in building the speech tokenizer with 65536 codewords. First, the vanilla VQ still performs the worst, and we notice that replacing dead codewords degrades the training stability in this large codebook. Both RVQ and FSQ can address ``index collapse" effectively. They apply the encoding space with strong constraints (residual embedding and 1-dim bottleneck feature) to pursue high codebook usage and perplexity. However, these constraints also limit the encoder in extracting effective information for feature reconstruction. Conversely, the proposed PQ-VAE with larger slackness achieves the best reconstruction quality with slightly lower codebook usage and perplexity than FSQ. It better meets the requirements of a high-quality speech tokenizer, i.e., improving codebook quality to reduce reconstruction loss.

\begin{table}[htp]
\centering
\caption{The evaluation results on test-large of autoencoders with $65536$ codewords and different quantization approaches.}
\label{tab:fsq}
\begin{tabular}{l|ccc}
\toprule
Method & Usage & Perplexity & RMSE \\ \midrule
VQ  & 2195 & 1597 & 7.31 \\
RVQ & 61264 & 29246 & 7.05 \\
FSQ  & \textbf{62500} & \textbf{34720} & 7.04 \\ \midrule
PQ-VAE & 60604 & 29531 & \textbf{6.93} \\ \bottomrule
\end{tabular}
\end{table}

\subsection{Application in TTS}

Finally, we enhance speech tokenizers with HuBERT features to investigate their performance in LLM-based TTS. As shown in Table \ref{tab:tts}, VQ-8192, i.e. the vanilla VQ-VAE with 8192 codewords, still suffers from ``index collapse", providing only 700 valid codewords, leading to lower intelligibility in TTS. PQ-8192 addressing this issue shows higher codebook quality and lower reconstruction loss, achieving better TTS performance in naturalness, intelligibility, and speaker similarity. Then, a larger speech tokenizer, PQ-32768, further improves TTS quality. The large codebook with richer information enables TTS to produce clearer and more precise pronunciation with fewer mispronunciation issues. It validates that a large-codebook speech tokenizer is important for LLM-TTS, and the proposed PQ-VAE makes it achievable.


\begin{table}[htp]
\centering
\caption{The evaluation results of different speech tokenizers on feature reconstruction and TTS, where VQ-8192 denotes VQ-VAE with 8192 codewords, and so on.}
\label{tab:tts}
\begin{tabular}{cc|cccc}
\toprule
\multicolumn{2}{c|}{Model} & VQ-8192 & PQ-8192 & PQ-32768 \\ \midrule
\multicolumn{2}{c|}{Usage} & 699 & 8171 & \textbf{32731} \\
\multicolumn{2}{c|}{Perplexity} & 630 & 5958 & \textbf{23518} \\ 
\multicolumn{2}{c|}{RMSE} & 3.63 & 3.46 & \textbf{3.39} \\ \midrule
\multicolumn{1}{c|}{\multirow{3}{*}{TTS}} & NISQA $\uparrow$ & 3.99 & 4.07 & \textbf{4.20} \\
\multicolumn{1}{c|}{} & CER (\%) $\downarrow$ & 11.34 & 8.78 & \textbf{7.05} \\
\multicolumn{1}{c|}{} & SS $\downarrow$ & 0.1286 & 0.1222 & \textbf{0.1214} \\ \bottomrule 
\end{tabular}
\end{table}

\section{Conclusion}

This paper proposes PQ-VAE that utilizes product quantization with multiple small codebooks to build a large-codebook speech tokenizer. Besides, a dual-decoding training strategy is applied to ensure all VQ subspaces are utilized well. The experimental results show that PQ-VAE effectively solves ``index collapse" well and is enhanced by the proposed dual-decoding training strategy, keeping higher codebook usage for 65536 codewords. PQ is also validated as a better multi-codebook VQ approach for large-codebook speech tokenizers, outperforming RVQ and FSQ regarding codebook quality and reconstruction quality. Finally, the performance of PQ-VAE in TTS further demonstrates that a high-quality large-codebook speech tokenizer is crucial for LLM-based speech generation.

\bibliographystyle{IEEEbib}
\bibliography{mybib}

\begin{thebibliography}{10}

\bibitem{brown2020language}
Tom Brown, Benjamin Mann, Nick Ryder, Melanie Subbiah, Jared~D Kaplan, Prafulla Dhariwal, Arvind Neelakantan, Pranav Shyam, Girish Sastry, Amanda Askell, et~al.,
\newblock ``Language models are few-shot learners,''
\newblock {\em Advances in neural information processing systems}, vol. 33, pp. 1877--1901, 2020.

\bibitem{openai2023gpt4}
OpenAI,
\newblock ``Gpt-4 technical report,'' 2023.

\bibitem{touvron2023llama2}
Hugo Touvron, Louis Martin, Kevin Stone, Peter Albert, Amjad Almahairi, Yasmine Babaei, Nikolay Bashlykov, Soumya Batra, Prajjwal Bhargava, Shruti Bhosale, et~al.,
\newblock ``Llama 2: Open foundation and fine-tuned chat models,''
\newblock {\em arXiv preprint arXiv:2307.09288}, 2023.

\bibitem{kaplan2020scaling}
Jared Kaplan, Sam McCandlish, Tom Henighan, Tom~B Brown, Benjamin Chess, Rewon Child, Scott Gray, Alec Radford, Jeffrey Wu, and Dario Amodei,
\newblock ``Scaling laws for neural language models,''
\newblock {\em arXiv preprint arXiv:2001.08361}, 2020.

\bibitem{hassid2024textually}
Michael Hassid, Tal Remez, Tu~Anh Nguyen, Itai Gat, Alexis Conneau, Felix Kreuk, Jade Copet, Alexandre Defossez, Gabriel Synnaeve, Emmanuel Dupoux, et~al.,
\newblock ``Textually pretrained speech language models,''
\newblock {\em Advances in Neural Information Processing Systems}, vol. 36, 2024.

\bibitem{wu2023speechgen}
Haibin Wu, Kai-Wei Chang, Yuan-Kuei Wu, and Hung-yi Lee,
\newblock ``Speechgen: Unlocking the generative power of speech language models with prompts,''
\newblock {\em arXiv preprint arXiv:2306.02207}, 2023.

\bibitem{betker2023better}
James Betker,
\newblock ``Better speech synthesis through scaling,''
\newblock {\em arXiv preprint arXiv:2305.07243}, 2023.

\bibitem{lajszczak2024base}
Mateusz {\L}ajszczak, Guillermo C{\'a}mbara, Yang Li, Fatih Beyhan, Arent van Korlaar, Fan Yang, Arnaud Joly, {\'A}lvaro Mart{\'\i}n-Cortinas, Ammar Abbas, Adam Michalski, et~al.,
\newblock ``Base tts: Lessons from building a billion-parameter text-to-speech model on 100k hours of data,''
\newblock {\em arXiv preprint arXiv:2402.08093}, 2024.

\bibitem{wang2023lm}
Zhichao Wang, Yuanzhe Chen, Lei Xie, Qiao Tian, and Yuping Wang,
\newblock ``Lm-vc: Zero-shot voice conversion via speech generation based on language models,''
\newblock {\em arXiv preprint arXiv:2306.10521}, 2023.

\bibitem{kuan2023towards}
Chun-Yi Kuan, Chen-An Li, Tsu-Yuan Hsu, Tse-Yang Lin, Ho-Lam Chung, Kai-Wei Chang, Shuo-Yiin Chang, and Hung-yi Lee,
\newblock ``Towards general-purpose text-instruction-guided voice conversion,''
\newblock in {\em 2023 IEEE Automatic Speech Recognition and Understanding Workshop (ASRU)}. IEEE, 2023, pp. 1--8.

\bibitem{huang2023speech}
Zhichao Huang, Rong Ye, Tom Ko, Qianqian Dong, Shanbo Cheng, Mingxuan Wang, and Hang Li,
\newblock ``Speech translation with large language models: An industrial practice,''
\newblock {\em arXiv preprint arXiv:2312.13585}, 2023.

\bibitem{vqvae}
Aaron van~den Oord, Oriol Vinyals, and koray kavukcuoglu,
\newblock ``Neural discrete representation learning,''
\newblock in {\em Advances in Neural Information Processing Systems}, I.~Guyon, U.~Von Luxburg, S.~Bengio, H.~Wallach, R.~Fergus, S.~Vishwanathan, and R.~Garnett, Eds. 2017, vol.~30, Curran Associates, Inc.

\bibitem{garbacea2019low}
Cristina G{\^a}rbacea, A{\"a}ron van~den Oord, Yazhe Li, Felicia~SC Lim, Alejandro Luebs, Oriol Vinyals, and Thomas~C Walters,
\newblock ``{Low bit-rate speech coding with VQ-VAE and a WaveNet decoder},''
\newblock in {\em Proc. ICASSP}. IEEE, 2019, pp. 735--739.

\bibitem{wang2021vqmivc}
Disong Wang, Liqun Deng, Yu~Ting Yeung, Xiao Chen, Xunying Liu, and Helen Meng,
\newblock ``Vqmivc: Vector quantization and mutual information-based unsupervised speech representation disentanglement for one-shot voice conversion,''
\newblock {\em arXiv preprint arXiv:2106.10132}, 2021.

\bibitem{Du2022VQTTSHT}
Chenpeng Du, Yiwei Guo, Xie Chen, and K.~Yu,
\newblock ``{VQTTS: High-fidelity text-to-speech synthesis with self-supervised VQ acoustic feature},''
\newblock in {\em Proc. Interspeech}, 2022.

\bibitem{huang2023repcodec}
Zhichao Huang, Chutong Meng, and Tom Ko,
\newblock ``Repcodec: A speech representation codec for speech tokenization,''
\newblock {\em arXiv preprint arXiv:2309.00169}, 2023.

\bibitem{huh2023improvedvqste}
Minyoung Huh, Brian Cheung, Pulkit Agrawal, and Phillip Isola,
\newblock ``Straightening out the straight-through estimator: Overcoming optimization challenges in vector quantized networks,''
\newblock in {\em International Conference on Machine Learning}. PMLR, 2023.

\bibitem{guo2023msmc}
Haohan Guo, Fenglong Xie, Xixin Wu, Frank~K Soong, and Helen MengFellow,
\newblock ``Msmc-tts: Multi-stage multi-codebook vq-vae based neural tts,''
\newblock {\em IEEE/ACM Transactions on Audio, Speech, and Language Processing}, 2023.

\bibitem{DBLP:journals/corr/abs-2210-13438}
Alexandre D{\'{e}}fossez, Jade Copet, Gabriel Synnaeve, and Yossi Adi,
\newblock ``High fidelity neural audio compression,''
\newblock {\em CoRR}, vol. abs/2210.13438, 2022.

\bibitem{wang2023neural}
Chengyi Wang, Sanyuan Chen, Yu~Wu, Ziqiang Zhang, Long Zhou, Shujie Liu, Zhuo Chen, Yanqing Liu, Huaming Wang, Jinyu Li, et~al.,
\newblock ``Neural codec language models are zero-shot text to speech synthesizers,''
\newblock {\em arXiv preprint arXiv:2301.02111}, 2023.

\bibitem{zhang2023speechtokenizer}
Xin Zhang, Dong Zhang, Shimin Li, Yaqian Zhou, and Xipeng Qiu,
\newblock ``Speechtokenizer: Unified speech tokenizer for speech large language models,''
\newblock {\em arXiv preprint arXiv:2308.16692}, 2023.

\bibitem{jegou2010product}
Herve Jegou, Matthijs Douze, and Cordelia Schmid,
\newblock ``{Product quantization for nearest neighbor search},''
\newblock {\em IEEE transactions on pattern analysis and machine intelligence}, vol. 33, no. 1, pp. 117--128, 2010.

\bibitem{Oord2017NeuralDR}
A{\"a}ron van~den Oord, Oriol Vinyals, and Koray Kavukcuoglu,
\newblock ``{Neural discrete representation learning},''
\newblock in {\em Proc. NeurIPS}, 2017.

\bibitem{zheng2023online}
Chuanxia Zheng and Andrea Vedaldi,
\newblock ``Online clustered codebook,''
\newblock in {\em Proceedings of the IEEE/CVF International Conference on Computer Vision}, 2023, pp. 22798--22807.

\bibitem{zhang2022wenetspeech}
Binbin Zhang, Hang Lv, Pengcheng Guo, Qijie Shao, Chao Yang, Lei Xie, Xin Xu, Hui Bu, Xiaoyu Chen, Chenchen Zeng, et~al.,
\newblock ``Wenetspeech: A 10000+ hours multi-domain mandarin corpus for speech recognition,''
\newblock in {\em ICASSP 2022-2022 IEEE International Conference on Acoustics, Speech and Signal Processing (ICASSP)}. IEEE, 2022, pp. 6182--6186.

\bibitem{loshchilov2017decoupled}
Ilya Loshchilov and Frank Hutter,
\newblock ``Decoupled weight decay regularization,''
\newblock {\em arXiv preprint arXiv:1711.05101}, 2017.

\bibitem{hsu2021hubert}
Wei-Ning Hsu, Benjamin Bolte, Yao-Hung~Hubert Tsai, Kushal Lakhotia, Ruslan Salakhutdinov, and Abdelrahman Mohamed,
\newblock ``Hubert: Self-supervised speech representation learning by masked prediction of hidden units,''
\newblock {\em IEEE/ACM Transactions on Audio, Speech, and Language Processing}, vol. 29, pp. 3451--3460, 2021.

\bibitem{guo2023qs}
Haohan Guo, Fenglong Xie, Jiawen Kang, Yujia Xiao, Xixin Wu, and Helen Meng,
\newblock ``Qs-tts: Towards semi-supervised text-to-speech synthesis via vector-quantized self-supervised speech representation learning,''
\newblock {\em arXiv preprint arXiv:2309.00126}, 2023.

\bibitem{tortoise}
James Betker,
\newblock ``Better speech synthesis through scaling,''
\newblock {\em arXiv preprint arXiv:2305.07243}, 2023.

\bibitem{bigvgan}
{Sang-gil} Lee, Wei Ping, Boris Ginsburg, Bryan Catanzaro, and Sungroh Yoon,
\newblock ``Big{VGAN}: A universal neural vocoder with large-scale training,''
\newblock in {\em The Eleventh International Conference on Learning Representations}, 2023.

\bibitem{mittag2021deep}
Gabriel Mittag and Sebastian M{\"o}ller,
\newblock ``Deep learning based assessment of synthetic speech naturalness,''
\newblock {\em arXiv preprint arXiv:2104.11673}, 2021.

\bibitem{chen2010approximate}
Yongjian Chen, Tao Guan, and Cheng Wang,
\newblock ``Approximate nearest neighbor search by residual vector quantization,''
\newblock {\em Sensors}, vol. 10, no. 12, pp. 11259--11273, 2010.

\bibitem{mentzer2023finite}
Fabian Mentzer, David Minnen, Eirikur Agustsson, and Michael Tschannen,
\newblock ``Finite scalar quantization: Vq-vae made simple,''
\newblock {\em arXiv preprint arXiv:2309.15505}, 2023.

\end{thebibliography}

\end{document}